\documentclass[twocolumn,aps,prd,eqsecnum,nofootinbib,showpacs]{revtex4}
\usepackage{verbatim}
\usepackage{amssymb}
\usepackage{amsmath}
\usepackage{amsfonts}
\usepackage{bm}

\begin{document}

\title{Covariant formulation of the post-1-Newtonian
       approximation to General Relativity}

\author{Wolfgang Tichy}
\affiliation{Department of Physics, Florida Atlantic University,
             Boca Raton, FL  33431}

\author{\'Eanna \'E.\ Flanagan}
\affiliation{Center for Radiophysics and Space Research,
             Cornell University, Ithaca, NY 14853}

%
%
\newcount\hh
\newcount\mm
\mm=\time
\hh=\time
\divide\hh by 60
\divide\mm by 60
\multiply\mm by 60
\mm=-\mm
\advance\mm by \time
\def\hhmm{\number\hh:\ifnum\mm<10{}0\fi\number\mm}


\date{draft of June 22, 2011; printed \today{} at \hhmm}

\begin{abstract}
We derive a coordinate-independent formulation of the post-1-Newtonian
approximation to general relativity. This formulation is a
generalization of the Newton-Cartan
geometric formulation of Newtonian gravity.
It involves several fields and a connection, but no spacetime metric
at the fundamental level. We show that the usual
coordinate-dependent equations of post-Newtonian gravity are recovered
when one specializes to asymptotically flat spacetimes
and to appropriate classes of coordinates.
\end{abstract}


\maketitle

\def\be{\begin{equation}}
\def\ee{\end{equation}}
\def\bea{\begin{eqnarray}}
\def\eea{\end{eqnarray}}
\def\nn{\nonumber}
\newcommand{\bes}{\begin{subequations}}
\newcommand{\ees}{\end{subequations}}

\section{Introduction and Summary}

\subsection{Background and Motivation}

The weak field, slow motion
approximation to general relativity, also called the post-Newtonian
approximation, consists of
expanding in the small
parameters $v^2/c^2$ and $\Phi/c^2$, where $v$ is a
typical velocity
of the system under consideration, $\Phi$ is the Newtonian potential,
and $c$ is the speed of light.
At the leading order Newton's theory is recovered, and higher order
corrections are called post-1-Newtonian corrections, post-2-Newtonian
corrections and so on.
This approximation scheme is very useful in astrophysical applications
and is very well developed.  Reviews can be found in Ref.\ \cite{blanchet}
and in the book by Will \cite{Will-book}, and a historical review can be found in Ref.\ \cite{1987thyg.book..128D}.

There are two types of equations that arise in post-Newtonian
theory.  The first are continuum equations of motion, for example for
gravity coupled to a perfect fluid, for which one obtains
generalizations of the equations of Newtonian hydrodynamics.  Such
continuum equations have been used extensively in numerical
simulations (although fully relativistic simulations are now the
state of the art \cite{2010CQGra..27k4002D}).
The second type of equations are ``point particle equations'', which
describe the motions of bodies whose sizes are small compared to their
separations.  Such point particle equations (and extensions to include
spins) can be derived from the underlying continuum theory by a
variety of methods.  Currently the equations of motion for two point
particles are known up to post-3.5-Newtonian order, see, for example
Ref.\ \cite{2009arXiv0907.3596B} and references therein.
In this paper we shall be concerned only with continuum equations.

Over the years, the foundations of the Newtonian and post-Newtonian
approximations have been studied in detail and with considerable
mathematical rigor
by a number of researchers.  Futamase and Schutz \cite{Futamase-S}
have shown that the various orders of post-Newtonian approximation are
asymptotic approximations to fully relativistic solutions.
Frittelli and Reula \cite{1994CMaPh.166..221F} showed that, given a
solution of the equations of Newtonian gravity, there exists a
one-parameter family of exact relativistic solutions which for a
finite
amount of time are close to the Newtonian solution.
Rendall \cite{Rendall} gave a mathematically rigorous derivation of
both the Newtonian and post-Newtonian approximations from a precise
set of axioms.

However, there remains one aspect of post-Newtonian theory which has
not been fully explored: there is as yet no covariant version of the theory.
Usually, in order to find the equations of post-Newtonian theory,
one first introduces specializations of the coordinate system (or gauge), and
then Einstein's equations are expanded order by order.
The gauge specializations are chosen to simplify the resulting equations.
For example, at post-1-Newtonian order, the harmonic gauge condition
and the so-called standard post-Newtonian gauge condition
\cite{Damour} are often used.
In each of these gauges the post-Newtonian equations take a different form.
The situation is similar to knowing the laws of electromagnetism only
in a handful of gauges (e.g. the Lorentz and the Coulomb gauge),
without knowing the underlying gauge independent equations.

This lack of covariance of post-Newtonian theory as currently formulated
has several disadvantages:
\begin{itemize}

\item If one is attempting to compare two different calculations, it
    is often helpful to identify and compute gauge-invariant quantities.
    Generally, such quantities are easier to identify starting from a covariant
    formulation of the theory.

\item In attempting to developing intuition from the post-Newtonian
  equations, it can be difficult to sort out which aspects of the
  equations contain the actual gauge-independent physics and which
  aspects are gauge
  dependent.  For example, there is a well-known analogy between the
  equations of post-1-Newtonian theory in certain gauges and those of
  electromagnetism.  In this case the analogy with electrostatics and
  magnetostatics is
  physical, but the additional aspects of the analogy concerning
  magnetodynamics are gauge.\footnote{If the analogy with
    electromagnetism were complete there would be radiation in the
    post-1-Newtonian theory.}

\item At post-1-Newtonian order, there is a well-developed framework
 for celestial mechanics that describes $N$ interacting, deformable
 bodies \cite{Damour,1992PhRvD..45.1017D,1993PhRvD..47.3124D,RF}.  This
 framework is quite complicated, involving separate coordinate systems
 for each body as well as a global coordinate system, together with
 a set of fields associated with each of the coordinate systems.
A covariant formulation might simplify some aspects of this
framework.

\end{itemize}

\subsection{Covariant post-Newtonian theory}

The purpose of this paper is to derive a coordinate independent
formulation of post-1-Newtonian theory.
We will derive a fully covariant set of equations, involving a number
of tensor fields and a connection, which reduce to the standard
post-Newtonian equations in specific coordinate systems for
asymptotically flat spacetimes.
In the case of Newton's theory, such a geometric formulation
has already been found by Cartan and others
\cite{Cartan0,Cartan1,Friedrichs,Trautman,MTW}, and is called
Newton-Cartan theory.
Building on earlier work of Dautcourt \cite{Dautcourt0,Dautcourt}, our
derivation will reproduce Newton-Cartan theory at
leading order, and at the next order will give a covariant version of
post-1-Newtonian theory which we call ``post-Newton-Cartan theory''.

We will actually derive two different versions of post-Newton-Cartan
theory.  The first version, which we
call perturbative post-Newton-Cartan theory, allows one to compute the
leading order corrections to a solution of Newton-Cartan theory.  It
is a unique theory, in which post-Newtonian corrections to Newtonian
quantities are treated as independent variables to be solved for.
The second version, which we call combined post-Newton-Cartan theory,
combines some of the Newtonian and post-Newtonian variables together.  It is
more economical and convenient to use than the first version, because it has fewer
variables and fewer equations.  Solutions of this theory will be
accurate to post-1-Newtonian order, but will in addition contain
post-2-Newtonian, post-3-Newtonian etc. pieces.  The combined theory is not
unique; different choices could be made to define different theories
whose solutions differ at post-2-Newtonian and higher orders.
We note that standard, coordinate-specific post-Newtonian theory also
comes in ``perturbative'' and ``combined'' versions \cite{Futamase-S}.

\subsection{Results}

We now turn to a description of our results.  We start by reviewing
Newton-Cartan theory, then we describe the two versions of
post-Newton-Cartan theory.

\subsubsection{Newton-Cartan theory}

\begin{table}
\caption{Fields and equations of Newton-Cartan theory.
\label{NC-table}}
\begin{tabular}{l|l}
\hline\hline
Field          & Description    \\
\hline
$h^{ab}$ &spatial metric, $(0,+,+,+)$  \\
$t_a$ & time one-form  \\
$D_a$ & symmetric connection  \\
${\cal T}^{ab}$ & matter stress-energy tensor \\
\hline\hline
Equation       & Description \\
\hline
$h^{ab} t_b= 0$ & Orthogonality   \\
$D_a h^{bc}=0$ &Compatibility with connection \\
$D_a t_b =0$ & Compatibility with connection \\
$h_{\mathstrut}^{e[a}R_{e(bc)}^{\ \ \  \ \ d]}=0$ & Trautman condition\\
$R_{ab}=4\pi t_a t_b t_c t_d {\cal T}^{cd}$ & Field equation\\
$D_a {\cal T}^{ab}=0$ & Stress-energy conservation \\
$R_{abc}^{\ \ \ d}=R_{abc}^{\ \ \ d}(D_e)$ & definition of Riemann \\
\end{tabular}
\end{table}

\begin{table*}
\caption{Additional fields and equations of perturbative post-Newton-Cartan theory.
\label{PNC-table}}
\begin{tabular}{l|l}
\hline\hline
Field          & Description    \\
\hline
$k^{ab}$ & contravariant metric perturbation  \\
$p_{ab}$ & covariant metric perturbation  \\
$\Delta^c_{\ ab}$ & connection perturbation \\
${\cal S}^{ab}$ & matter stress-energy perturbation \\
\hline\hline
Equation       & Description \\
\hline
$h^{ab} p_{bc}-k^{ab}t_b t_c=\delta^{a}_c$
& Orthogonality   \\
$D_a k^{bc}+\Delta^{b}_{\ ad}h^{dc}
           +\Delta^{c}_{\ ad}h^{bd} =0$
&Compatibility with connection \\
$D_a p_{bc}+\Delta^{d}_{\ ab}t_{d} t_c
           +\Delta^{d}_{\ ac}t_{b} t_d  =0$
 & Compatibility with connection \\
$k_{\mathstrut}^{e[a}  R_{e(bc)}^{\ \ \ \ \ d]}
- h_{\mathstrut}^{e[a} D^{\mathstrut}_e \Delta^{d]}_{\ \,bc} +
h_{\mathstrut}^{e[a} D^{\mathstrut}_{(b} \Delta^{d]}_{\ \,c)e}=0$
 & Trautman condition\\
$- D_a \Delta^{b}_{\ bc} + D_b \Delta^{b}_{\ ac}=
4 \pi \big[t_a t_b t_{c}t_{d}{\cal S}^{cd}
  -4t_{c}t_{(a}p_{b)d}{\cal T}^{cd}
  +p_{cd}{\cal T}^{cd}t_{a}t_{b}
  +t_{c}t_{d}{\cal T}^{cd}p_{ab}\big] $
& Field equation\\
$D_a {\cal S}^{ab}+\Delta^{a}_{\ ac}{\cal T}^{cb}
                +\Delta^{b}_{\ ac} {\cal T}^{ac}=0 $
 & Stress-energy conservation \\
\end{tabular}
\end{table*}

The variables of Newton-Cartan theory are a one-form $t_a$, a
symmetric contravariant tensor field $h^{ab}$ with signature $(0,+,+,+)$,
and a torsion-free connection $D_a$.  Matter is described by a
symmetric contravariant stress-energy tensor ${\cal T}^{ab}$.  The fields
$t_a$ and $h^{ab}$ are nondynamical, background fields, while $D_a$
and ${\cal T}^{ab}$ are dynamical.

The equations of the theory are the orthogonality condition
\be
h^{ab} t_b=0,
\label{orthogonality}
\ee
the compatibility of the fields $h^{ab}$ and $t_a$ with the connection,
\bes
\label{compat}
\bea
\label{compat-upper_0}
D_a h^{bc}&=&0 ,\\
D_a t_b &=&0,
\label{compat-lower_-2}
\eea
\ees
and the Trautman condition
\be
h_{\mathstrut}^{f[a} R_{f(bc)}^{\ \ \ \ \ d]}=0,
\label{Trautman0}
\ee
where $R_{abc}^{\ \ \ \,d}$ is the Riemann tensor associated with the
connection $D_a$.  In addition we have the field equation
\be
R_{ab} = 4 \pi t_a t_b t_c t_d {\cal T}^{cd},
\label{field-eq_0}
\ee
where $R_{ab} = R_{acb}^{\ \ \ \,c}$,
and the stress-energy conservation equation
\be
D_a {\cal T}^{ab} = 0.
\label{matter-eq_4}
\ee
These fields and equations are summarized in Table \ref{NC-table}.
In Sec.\ \ref{sec:derive1} below we review the derivation of this
Newton-Cartan
theory from general relativity, and in Sec.\ \ref{subsec-Newt-coord}
we review how
Newton-Cartan theory reduces to standard Newtonian gravity in
appropriate circumstances and in appropriate coordinate systems.

\begin{table*}
\caption{Fields and equations of combined post-Newton-Cartan theory.
\label{PNCC-table}}
\begin{tabular}{l|l}
\hline\hline
Field          & Description    \\
\hline
${\hat h}^{ab}$ &spatial metric, $(0,+,+,+)$  \\
${\hat t}_a$ & time one-form  \\
${\hat k}^{ab}$ & contravariant metric perturbation  \\
${\hat p}_{ab}$ & covariant metric perturbation  \\
${\hat D}_a$ & symmetric connection  \\
${\hat {\cal T}}^{ab}$ & matter stress-energy tensor \\

\hline\hline
Equation       & Description \\
\hline
${\hat h}^{ab} {\hat t}_b= 0$ & Newtonian orthogonality   \\
${\hat h}^{ab} {\hat p}_{bc}-{\hat k}^{ab}{\hat t}_b {\hat t}_c=\delta^{a}_c$
& post-Newtonian orthogonality   \\
${\hat D}_a ( {\hat h}^{bc} + {\hat k}^{bc})=0$ &Compatibility with connection \\
${\hat D}_a (- {\hat t}_b {\hat t}_c + {\hat p}_{bc})=0$ & Compatibility with connection \\
$({\hat h}_{\mathstrut}^{e[a} + {\hat k}_{\mathstrut}^{e[a})
  R_{e(bc)}^{\ \ \  \ \ d]}({\hat D}_e)=0$ & Trautman condition\\
$R_{ab}({\hat D}_c)=
4 \pi \big[{\hat t}_a {\hat t}_b {\hat t}_{c}{\hat t}_{d}{\hat {\cal T}}^{cd}
  -4 {\hat t}_{c} {\hat t}_{(a}{\hat p}_{b)d}{\hat {\cal T}}^{cd}
  +{\hat p}_{cd}{\hat {\cal T}}^{cd}{\hat t}_{a}{\hat t}_{b}
  +{\hat t}_{c}{\hat t}_{d}{\hat {\cal T}}^{cd}{\hat p}_{ab}\big] $
& Field equation\\
${\hat D}_a {\hat {\cal T}}^{ab}=0$ & Stress-energy conservation \\
\end{tabular}
\end{table*}

\subsubsection{Perturbative post-Newton-Cartan theory}

Turn, now to the perturbative post-Newton-Cartan theory.
This theory
contains the Newtonian fields $h^{ab}$, $t_a$, $D_a$ and ${\cal T}^{ab}$, and
in addition four new fields: a symmetric, contravariant tensor
$k^{ab}$, a symmetric, covariant tensor $p_{ab}$, a
perturbation $\Delta^c_{\ ab}$ to the connection, and a
perturbation ${\cal S}^{ab}$ to the matter stress-energy tensor.

The equations of the theory are the six Newtonian equations
(\ref{orthogonality}) -- (\ref{matter-eq_4}), together with six
post-Newtonian equations:
(i) the orthogonality condition
\be
h^{ab} p_{bc}-k^{ab}t_b t_c=\delta^{a}_c;
\label{inv-metric0}
\ee
(ii) the compatibility of $k^{ab}$ with the connection,
\be
D_a k^{bc}+\Delta^{b}_{\ ad}h^{dc}
           +\Delta^{c}_{\ ad}h^{bd} =0;
\label{compat-upper_2}
\ee
(iii) the compatibility of $p_{ab}$ with the connection,
\be
D_a p_{bc}+\Delta^{d}_{\ ab}t_{d} t_c
           +\Delta^{d}_{\ ac}t_{b} t_d  =0;
\label{compat-lower_0}
\ee
(iv) the post-Newtonian Trautman condition
\be
k_{\mathstrut}^{e[a}  R_{e(bc)}^{\ \ \ \ \ d]}
- h_{\mathstrut}^{e[a} D^{\mathstrut}_e \Delta^{d]}_{\ \,bc} +
h_{\mathstrut}^{e[a} D^{\mathstrut}_{(b} \Delta^{d]}_{\ \,c)e}=0;
\label{Trautman2}
\ee
(v) the field equation
\begin{eqnarray}
- D_a \Delta^{b}_{\ bc} + D_b \Delta^{b}_{\ ac}&=&
4 \pi \big[t_a t_b t_{c}t_{d}{\cal S}^{cd}
  -4t_{c}t_{(a}p_{b)d}{\cal T}^{cd} \nonumber \\
&&  +p_{cd}{\cal T}^{cd}t_{a}t_{b}
  +t_{c}t_{d}{\cal T}^{cd}p_{ab}\big]; \nonumber \\
\label{field-eq_2}
\end{eqnarray}
and (vi) the stress-energy conservation equation
\be
D_a {\cal S}^{ab}+\Delta^{a}_{\ ac}{\cal T}^{cb}
                +\Delta^{b}_{\ ac} {\cal T}^{ac}=0.
\label{matter-eq_6}
\ee
These fields and equations are summarized in Table \ref{PNC-table}.
In Sec.\ \ref{sec:derive1} below we review the derivation of this perturbative
post-Newton-Cartan
theory from general relativity, and in Sec.\ \ref{sec-PN-fields} we review how
it reduces to standard post-1-Newtonian gravity in
appropriate circumstances and in appropriate coordinate systems.

\subsubsection{Combined post-Newton-Cartan theory}

The combined post-Newton-Cartan theory has aspects of both the
Newton-Cartan and the perturbative post-Newton-Cartan theories.
The independent variables are a one-form ${\hat t}_a$, a
symmetric contravariant tensor field ${\hat h}^{ab}$ with signature
$(0,+,+,+)$, a symmetric, contravariant tensor
${\hat k}^{ab}$, and a symmetric, covariant tensor ${\hat p}_{ab}$.
There is also a torsion-free connection ${\hat D}_a$, which is defined
accurate to post-1-Newtonian order.  Matter is described by a
symmetric contravariant stress-energy tensor ${\hat {\cal T}}^{ab}$,
which is also defined accurate to post-1-Newtonian order.
The fields
${\hat t}_a$ and ${\hat h}^{ab}$ are nondynamical, background fields,
while the remaining fields are dynamical.

The equations of the theory are: (i) the orthogonality conditions
\bes
\bea
\label{orthog0CPNC}
{\hat h}^{ab} {\hat t}_b&=&0, \\
{\hat h}^{ab} {\hat p}_{bc}-{\hat k}^{ab}{\hat t}_b {\hat t}_c&=&\delta^{a}_c;
\label{orthogonalityCPNC}
\eea
\ees
(ii) the connection compatibility conditions
\bes
\label{compatCPNC}
\bea
\label{compat-upper_0CPNC}
{\hat D}_a ( {\hat h}^{bc} + {\hat k}^{bc})&=&0, \\
{\hat D}_a (- {\hat t}_b {\hat t}_c + {\hat p}_{bc})&=&0;
\label{compat-lower_-2CPNC}
\eea
\ees
(iii) the Trautman condition
\be
({\hat h}_{\mathstrut}^{e[a} + {\hat k}_{\mathstrut}^{e[a})
  R_{e(bc)}^{\ \ \  \ \ d]}({\hat D}_e)=0;
\label{Trautman0CPNC}
\ee
(iv) the field equation
\bea
R_{ab}({\hat D}_c)&=&
4 \pi \big[{\hat t}_a {\hat t}_b {\hat t}_{c}{\hat t}_{d}{\hat {\cal T}}^{cd}
  -4 {\hat t}_{c} {\hat t}_{(a}{\hat p}_{b)d}{\hat {\cal T}}^{cd}
  \nonumber \\
&&  +{\hat p}_{cd}{\hat {\cal T}}^{cd}{\hat t}_{a}{\hat t}_{b}
  +{\hat t}_{c}{\hat t}_{d}{\hat {\cal T}}^{cd}{\hat p}_{ab}\big];
\label{field-eq_0CPNC}
\eea
and (v) the stress-energy conservation equation
\be
{\hat D}_a {\hat {\cal T}}^{ab} = 0.
\label{matter-eq_4CPNC}
\ee
These fields and equations are summarized in Table \ref{PNCC-table}.
The derivation of this combined post-Newton-Cartan theory from the
Newton-Cartan and post-Newton-Cartan theories is given in Sec.\
\ref{CPNC-derive} below.

\subsection{Organization of this paper}

In Sec.\ \ref{sec:foundations} we list the assumptions underlying
our derivation, and discuss the motivation for these assumptions.
Section \ref{sec-deriv} gives the derivation of the Newton-Cartan
theory and both versions of the post-Newton-Cartan theory, from these
assumptions.  In Sec.\ \ref{coord-PN} we show
that the equations of perturbative post-Newton-Cartan theory
reduce to the standard coordinate-specific equations of
post-Newtonian theory, under certain conditions and in certain
coordinate systems.  We conclude by summarizing our results and their
implications in Sec.\ \ref{Conclusion}.

Some technical and side issues are discussed in the Appendices.
Appendix \ref{ThirdAssumption} derives an alternative form of one of
our assumptions.  Appendix \ref{subsec-gauge-cov} derives the gauge
transformation properties of the fields of the perturbative
post-Newton-Cartan theory.  Finally Appendix \ref{perfect-fluid}
specializes the formalism, as an example, to perfect fluids.

\subsection{Notation}

Throughout we will use the metric and sign conventions of Misner,
Thorne and Wheeler \cite{MTW}. Furthermore we will use
Penrose's abstract index notation \cite{Wald}, with
indices $(a, b, \ldots)$ from the beginning of the Latin alphabet
denoting general tensors.  When we specialize to particular
coordinate systems, we will use indices $(\alpha,\beta, \ldots)$ from
the Greek alphabet to denote general components of tensors.
We will also use indices $(i, j, k, \ldots)$ from the
middle of the Latin alphabet to denote spatial
components of tensors, and an index $0$ to denote time components,
in particular coordinate systems.
For example, $u^a$ will denote a four-velocity vector, $u^{\mu}$ will
be the components of the four-velocity in some coordinate system,
and $u^i$ and $u^0$ will be the spatial and time components in this
coordinate system.  We will work with units in which $G=c=1$, except
for some special cases in Sec.\ \ref{sec:motivation}
where factors of $c$ and $G$ are explicitly included.
Finally, we will use the conventional definitions of the two order
symbols $O()$ and $o()$.

\section{Foundations and Assumptions}
\label{sec:foundations}

In this section we define and discuss the assumptions that we make in
order to derive
the Newton-Cartan and post-Newton-Cartan theories from general
relativity.  At Newtonian order we follow closely the treatments of
Dautcourt \cite{Dautcourt0,Dautcourt} and of Rendall \cite{Rendall}.

\subsection{Assumptions}
\label{sec:assumptions}

The starting point is to assume a one-parameter family
$g_{ab}(\varepsilon),T^{ab}(\varepsilon)$ of exact solutions of
Einstein's equations
\be
G^{ab}[g_{cd}(\varepsilon)] = 8 \pi T^{ab}(\varepsilon),
\label{field-eq00}
\ee
where $\varepsilon$ is a real parameter with $0 < \varepsilon <
\varepsilon_0$ for some $\varepsilon_0$.  We assume that the solutions
are smooth functions of spacetime and of $\varepsilon$ for
$\varepsilon >0$.  We do not assume the
existence of a solution at $\varepsilon=0$.

The key assumptions we make are:
\begin{enumerate}

\item The contravariant metric $g^{ab}(\varepsilon)$
can be expanded near $\varepsilon =0$ as
\begin{equation}
\label{metric-up}
g^{ab}(\varepsilon)=h^{ab}+\varepsilon k^{ab} + \varepsilon^2 j^{ab} +
\varepsilon^3 l^{ab}+o(\varepsilon^3),
\end{equation}
where $h^{ab}$, $k^{ab}$, $j^{ab}$ and $l^{ab}$ are $\varepsilon$-independent
symmetric tensor fields on spacetime.  Furthermore
$h^{ab}$ has signature
\begin{equation}
\label{h-signature}
h^{ab} \sim (0,+,+,+) ,
\end{equation}
and
$k^{ab} t_a t_b$ is everywhere nonzero, where $t_a$ is
the direction defined by $h^{ab} t_b=0$.

\item The matter stress-energy tensor can be expanded near
  $\varepsilon=0$ as
\begin{equation}
\label{matter_assumpt}
T^{ab}(\varepsilon)
=\varepsilon^2 {\cal T}^{ab} + \varepsilon^3 {\cal S}^{ab} +O(\varepsilon^4) ,
\end{equation}
where again ${\cal T}^{ab}$ and ${\cal S}^{ab}$ are
$\varepsilon$-independent tensor fields on spacetime.

\item The connection $\nabla_a$ associated with the metric
  $g_{ab}(\varepsilon)$ has a continuous limit $D_a$ as $\varepsilon
  \to 0$,
\be
\nabla_a = D_a + O(\varepsilon).
\label{connection-limit}
\ee

\end{enumerate}

\subsection{Motivation and Discussion}
\label{sec:motivation}

We now discuss the motivation for and properties of these
assumptions.

First, we note that, since the limiting contravariant metric $h^{ab}$
is degenerate, the limit as $\varepsilon\to0$ of the covariant metric
does not exist.  Therefore there is no limiting, background solution
of Einstein's equations at $\varepsilon=0$ in this framework, unlike
the situation for standard perturbation theory
\cite{Rendall,Dautcourt0,Dautcourt}.

Next, the assumptions are explicitly local and covariant.
In Sec.\ \ref{sec-deriv} below we will show that the Newton-Cartan and
post-Newton-Cartan theories can be derived from them in a
local and covariant manner.  Suppose now that one demands that the
assumptions apply only in a given, finite region of spacetime.  Then,
the Newton-Cartan and post-Newton-Cartan equations of Tables \ref{NC-table}
and \ref{PNC-table} will be satisfied in that region.
However, as is well known, it does {\it not} follow that the usual
equations of Newtonian gravity will be satisfied, since Newton-Cartan
theory contains more
local degrees of freedom than Newtonian gravity.  The physical reason
for this will be discussed in Sec.\ \ref{subsec-Newt-coord} below.
In order to obtain Newtonian gravity, it is necessary to assume an
asymptotically flat spacetime and to impose the assumptions (\ref{field-eq00})
-- (\ref{connection-limit}) throughout all of spacetime
\cite{Dautcourt0,Dautcourt,Rendall}.  We will find a similar situation
for the post-Newton-Cartan theory in Sec.\ \ref{coord-PN} below: it contains
more local degrees of
freedom than standard post-1-Newtonian general relativity, and reduces
to it only when the assumptions (\ref{field-eq00}) --
(\ref{connection-limit}) hold globally in an asymptotically flat
spacetime.

Consider now an isolated physical system that is characterized by some
mass scale ${\cal M}$, lengthscale ${\cal L}$, and timescale ${\cal
  T}$.  Then, from Newton's constant of gravitation $G$ and the speed
of light $c$ one can form two dimensionless parameters:
\be
{\hat c} \equiv \frac{c {\cal T}}{{\cal L}}, \ \ \ \
{\hat G} \equiv \frac{G {\cal M} {\cal T}^2}{{\cal L}^3}.
\ee
The Newtonian limit of general relativity is the limit ${\hat c} \to
\infty$ at fixed ${\hat G}$.

The first assumption, the expansion (\ref{metric-up}) of the
contravariant components of the metric, can now be motivated as
follows.  The spacetime metric should be close to the flat, Minkowski
metric in the Newtonian limit.  In system-adapted units where
${\cal T} = {\cal M} = {\cal L} = 1$, this is
$
ds^2 =-{\hat c}^2 dt^2 + dx^2 + dy^2 +dz^2.
$
Now identifying $\varepsilon = {\hat c}^{-2}$ gives
$$
g^{\mu\nu}= \mbox{diag}(-\varepsilon,1,1,1)=O(1)+O(\varepsilon) ,
$$
which satisfies assumption\ 1 and has signature
$(0,+,+,+)$ at
order $O(\varepsilon^0)$.  Corrections from the Newtonian potential do
not change this conclusion.

The second assumption, the expansion (\ref{matter_assumpt}) of the
stress-energy tensor, can be motivated similarly using dimensional
analysis.  In a general system of units Einstein's equation is
\be
G^{\alpha\beta} = \frac{8 \pi G}{c^4} T^{\alpha\beta},
\ee
where from dimensional analysis $T^{tt} \sim {\cal M} {\cal L}^{-3}$,
$T^{ti} \sim {\cal M} {\cal L}^{-2} {\cal T}^{-1}$, and $T^{ij} \sim
{\cal M} {\cal L}^{-1} {\cal T}^{-2}$.  In particular, all the
components of $T^{\alpha\beta}$ are independent of $c$ to leading
order.  If we now specialize to system-adapted units, all of the
components of $T^{\alpha\beta}$ are of order unity, and the right-hand
side of Einstein's equation is of order ${\hat G} / {\hat c}^4 \propto
\varepsilon^2$, since we are considering a limit in which ${\hat G}$ is
held fixed\footnote{If we replace the assumption
(\ref{matter_assumpt}) with a power series expansion of $T^{ab}$ that
starts with a term proportional to $\varepsilon^\nu$ with $\nu \ne 2$, we
would obtain a non-Newtonian limit of general relativity where ${\hat c} \to
\infty$ with ${\hat G} {\hat c}^{2 \nu -4}$ held fixed.
}.

Before discussing the third assumption, it is useful to consider the
gauge freedom present in the formalism.  It may appear that the
formalism so far is explicitly gauge invariant, since it is covariant
under general diffeomorphisms.  However, given a one-parameter family
of solutions $g_{ab}(\varepsilon), T^{ab}(\varepsilon)$ of Einstein's equations on a
manifold $M$, the gauge
freedom consists of a one-parameter family of diffeomorphisms
$\varphi_\varepsilon : M \to M$, which act on the solutions via
\be
g_{ab}(\varepsilon) \to \varphi_{\varepsilon\,*} \,g_{ab}(\varepsilon),
\ \ \ \ \ T^{ab}(\varepsilon) \to \varphi_{\varepsilon\,*} \,T^{ab}(\varepsilon).
\label{general-gauge-transformation}
\ee
Here $\varphi_{\varepsilon\,*}$ is the pullback mapping on tensor fields that is defined
in, for example, Appendix C of Ref.\ \cite{Wald}.
An important point is that, since we do not require the existence of a
solution at $\varepsilon =0$, there is no reason to require the
diffeomorphisms $\varphi_\varepsilon$ to have a well defined limit as
$\varepsilon \to 0$.  Thus, there are two subclasses of gauge
transformations:

\begin{itemize}

\item Transformations which we will call {\it regular}, consisting of
  smooth one-parameter families of diffeomorphisms which have a smooth
  limit as $\varepsilon \to 0$.  Such families can be parametrized in
  terms of a fixed, $\varepsilon$-independent diffeomorphism
  $\varphi_0$ and a set of vector fields $\xi^a_{(1)}$, $\xi^a_{(2)},
  \ldots$ via the expansion \cite{FW}
\be
\varphi_\varepsilon = \varphi_0 \circ {\cal D}_{{\vec
    \xi}_{(1)}}(\varepsilon) \circ {\cal D}_{{\vec
    \xi}_{(2)}}(\varepsilon^2) \circ \ldots,
\label{general-transf}
\ee
where for any vector field ${\vec \xi}$, ${\cal D}_{{\vec
    \xi}}(\varepsilon)$ is the diffeomorphism given by moving any
point $\varepsilon$ units along an integral curve of ${\vec \xi}$.

\item Transformations which we will call {\it irregular}, consisting of
  smooth one-parameter families $\varphi_\varepsilon$ of
  diffeomorphisms which have do not have a
  smooth limit as $\varepsilon \to 0$.

\end{itemize}

Our assumptions are explicitly covariant under regular gauge
transformations, as will be discussed in more detail in Appendix
\ref{subsec-gauge-cov}
below.  However, they are not covariant under irregular gauge
transformations.  For example, consider the prototypical Newtonian-order
metric that satisfies our assumptions (\ref{field-eq00})
-- (\ref{connection-limit}), namely
\bea
ds^2 &=& - \frac{1}{\varepsilon} \left[1 + 2 \varepsilon \Phi(t,x^i)
+ O(\varepsilon^2) \right] dt^2 \nonumber \\
&&
+ \left[ \delta_{ij} + O(\varepsilon) \right] dx^i dx^j,
\eea
where $\Phi$ is the Newtonian potential.
Under the irregular gauge transformation $t \to
\sqrt{\varepsilon} t$, $x^i \to x^i$, this metric is transformed into
\bea
ds^2 &=& - \left[1+ 2 \varepsilon \Phi(\sqrt{\varepsilon} t,x^i) + O(\varepsilon^2)\right]
dt^2 \nonumber \\
&& + \left[ \delta_{ij} + O(\varepsilon) \right] dx^i dx^j,
\eea
which does not satisfy our assumptions (\ref{field-eq00})
-- (\ref{connection-limit}).
Thus, our assumptions do entail a certain limited amount of gauge
specialization\footnote{
Because of the limited gauge dependence of our assumptions, it is not
{\it a priori} obvious that the fields $D_a$, $t_a$, $h^{ab}$,
$k^{ab}$, etc. that characterize the Newton-Cartan and
post-Newton-Cartan theories are physically unique.  More precisely,
suppose that we start with a one-parameter family of solutions
$g_{ab}(\varepsilon), T^{ab}(\varepsilon)$ which satisfies our
assumptions, and is thus characterized by a set of limiting fields
$D_a$, $t_a$, $h^{ab}$, $k^{ab}$ etc.  Now make a general (possibly irregular)
gauge transformation, to obtain a new one-parameter family
${\bar g}_{ab}(\varepsilon), {\bar T}^{ab}(\varepsilon)$ of solutions.
If this new family also satisfies our assumptions, then it will be
characterized by a new set of fields
${\bar D}_a$, ${\bar t}_a$, ${\bar h}^{ab}$, ${\bar k}^{ab}$.
We conjecture that in this case there must
exist a {\it regular} gauge transformation
of the form (\ref{general-transf}) relating the
two sets of fields (in the manner described in Appendix
\protect{\ref{subsec-gauge-cov}}), so
that
the fields are
unique in a physical sense.}, even though they are covariant under transformations
of the form (\ref{general-transf}).

We now turn to a discussion of the third assumption, the expansion
(\ref{connection-limit}) of the connection $\nabla_a$.
First, we remark that assumptions 1 and 2 are insufficient to
characterize the Newtonian limit.  For example, consider a
one-parameter family of
static vacuum spacetimes, of the form
\be
ds^2 = - e^{2 \alpha(x^k,\varepsilon)} dt^2 + h_{ij}(x^k,\varepsilon) dx^i dx^j,
\ee
which is smooth in $\varepsilon$ near $\varepsilon=0$.
This
one-parameter family of metrics, when written in terms of the
coordinates ${\bar t} = \sqrt{\varepsilon} t$ and $x^j$, satisfies our
assumptions 1 and 2, but is not of the type associated with the
Newtonian limit.  In particular components of the Riemann curvature
tensor will diverge in this example as $\varepsilon \to 0$.
Therefore some additional assumption like assumption 3 is
necessary.\footnote{Dautcourt \protect{\cite{Dautcourt}} shows that
assumption 3 follows from assumptions 1 and 2 when one makes additional assumptions about the global properties of the spacetime including asymptotic flatness.  However, we will not follow this route here, since we want to obtain a purely local derivation of the Newton-Cartan and post-Newton-Cartan theories.}
Our assumption 3 is actually slightly stronger than is necessary:
we show in Appendix
\ref{ThirdAssumption} that, whenever assumptions 1 and 2 hold
in a local region, and the Riemann tensor $R_{abc}^{\ \ \ \
  d}(\varepsilon)$ is finite as $\varepsilon \to 0$,
then there exists a (possibly irregular) gauge
transformation of the form (\ref{general-gauge-transformation}) such
that the transformed one-parameter family of
solutions satisfies assumptions 1, 2 and 3.

Finally we note that, if one wanted to go to higher post-Newtonian orders,
the assumptions used here to obtain a covariant
approximation scheme would
need to be modified.  First, to account for dissipative, radiative
effects, one would need
to introduce half-integer powers of $\varepsilon$ in the expansions.
These would first arise at order $O(\varepsilon^{5/2})$
in $g_{ab}$ and order $O(\varepsilon^{7/2})$ in $g^{ab}$.  Alternatively,
one could retain integer powers but make
the replacement $\varepsilon \to \varepsilon^2$ throughout.
Second, it is well known that solutions of the post-1-Newtonian field equations
are not good approximations to exact solutions at distances $
\gtrsim 1/\sqrt{\varepsilon}$.  That is, although they work well in the near
zone they break down in the local wave zone \cite{1980RvMP...52..299T}.
In order to find solutions which are good
approximations everywhere, one has to match
post-Newtonian solutions onto radiation zone post-Minkowskian solutions; see,
for example, Blanchet \cite{blanchet}.  However, the corresponding
corrections to the near zone gravitational fields and to the dynamics of the
bodies arises at post-2.5-Newtonian order, and will therefore not be
important for this paper.

\section{Derivation of Newton-Cartan and post-Newton-Cartan theories from General Relativity}
\label{sec-deriv}

In this section we derive the equations (\ref{orthogonality}) --
(\ref{matter-eq_4}) of Newton-Cartan theory,
(\ref{inv-metric0}) -- (\ref{matter-eq_6}) of
perturbative post-Newton-Cartan theory,
and (\ref{orthog0CPNC}) -- (\ref{matter-eq_4CPNC}) of combined post-Newton-Cartan theory,
from the assumptions discussed in Sec.\
\ref{sec:foundations} above.  The derivation will be local and
covariant\footnote{A similar approach to deriving a covariant
  post-Newtonian theory was undertaken by L. Gunnarsen (unpublished)
  at the University of Chicago in the 1980s.}.

We first note that it follows from assumption 1, together with an adjustment
of the
normalization of the one-form $t_a$ if necessary, that the covariant
components of the metric can be expanded as
\begin{equation}
\label{metric-down}
g_{ab}(\varepsilon)=- \frac{1}{\varepsilon} t_a t_b + p_{ab}
+ \varepsilon q_{ab} +o(\varepsilon),
\end{equation}
where $t_a$, $p_{ab}$ and $q_{ab}$
are $\varepsilon$-independent tensor fields on spacetime.
To see this, choose a basis of vector fields $e^a_{\hat \alpha}$ for
${\hat \alpha} = 0,1,2,3$ for which $h^{ab} = \delta^{{\hat i}{\hat
    j}} e^a_{{\hat i}} e^b_{{\hat j}}$.
Then by assumption we have $k^{{\hat 0}{\hat 0}}\ne0$, and
in fact $k^{{\hat 0}{\hat 0}}$ must be negative in order for
$g_{ab}(\varepsilon)$ to have signature $(-,+,+,+)$ for small
$\varepsilon$.  Now expanding and inverting on this basis
the expression
(\ref{metric-up}) for
the contravariant metric
yields an expression of the form
(\ref{metric-down}).

Next, in any coordinate system we can compute the coefficients
$\Gamma^\alpha_{\beta\gamma}(\varepsilon)$ of the connection by using
the expansions (\ref{metric-down}) and (\ref{metric-up}) of
the covariant and contravariant metrics.  This yields an expression of
the form
\be
\label{Gamma-schematic}
\Gamma^\alpha_{\beta\gamma}(\varepsilon) = O(\varepsilon^{-1}) +
O(\varepsilon^0) + O(\varepsilon) + o(\varepsilon),
\ee
where the first three terms can be computed explicitly from the fields
appearing in the metric expansions.  Now, from assumption 3 it follows
that the first term in (\ref{Gamma-schematic}) vanishes, and the
second term gives the coefficients of the Newtonian connection $D_a$
defined in Eq.\ (\ref{connection-limit}).  Therefore
we can write, for any one-form $w_a$,
\begin{equation}
\label{Da-def}
\nabla_a (\varepsilon) w_b= D_a w_b - \varepsilon \Delta^{c}_{\ ab}
w_c + o(\varepsilon),
\end{equation}
where $\Delta^c_{\ ab}$ is an $\varepsilon$-independent tensor field
which is symmetric in $a$ and $b$.  This quantity is
the post-Newtonian perturbation to the connection.
The $\varepsilon$ in brackets on the left hand side of Eq.\
(\ref{Da-def}) indicates the dependence of the derivative operator on
$\varepsilon$, not an application of the derivative operator to
$\varepsilon$.

\subsection{Newton-Cartan theory and perturbative post-Newton-Cartan theory}
\label{sec:derive1}

\subsubsection{Orthogonality conditions}

We start with the definition of the contravariant metric,
\be
g^{ab}(\varepsilon)g_{bc}(\varepsilon)=\delta^{a}_c,
\ee
and insert the expansions (\ref{metric-down}) and (\ref{metric-up}) of
the covariant and contravariant metrics.  At order
$O(\varepsilon^{-1})$ this yields the Newton-Cartan orthogonality
condition (\ref{orthogonality}), and at order $O(\varepsilon^0)$ it
yields the corresponding post-Newton-Cartan condition (\ref{inv-metric0}).

\subsubsection{Compatibility conditions}

Next, we insert the expansions (\ref{metric-down}) and (\ref{metric-up}) of
the covariant and contravariant metrics
and the expansion (\ref{Da-def}) of the connection
into the equations
\be
\nabla_a(\varepsilon) g_{bc}(\varepsilon) =  0, \ \ \
\nabla_a(\varepsilon) g^{bc}(\varepsilon) =  0.
\ee
At leading order, this yields the Newton-Cartan compatibility conditions
(\ref{compat-upper_0}) and (\ref{compat-lower_-2}), and at subleading
order one obtains the
post-Newton-Cartan compatibility conditions (\ref{compat-upper_2}) and (\ref{compat-lower_0}).

\subsubsection{Trautman conditions}

We next compute the Riemann tensor using the expansion (\ref{Da-def}) of the
connection.  This yields
\begin{eqnarray}
\label{Riemann}
R_{abc}^{\ \ \ d}[\nabla_e (\varepsilon)]&=&R_{abc}^{\ \ \ d}[D_e]
+ \varepsilon \big[
D_b \Delta^{d}_{\ ac}
-D_a \Delta^{d}_{\ bc}   \big]
\nonumber \\ &&
+ o(\varepsilon).
\end{eqnarray}
Here the first term on the right hand side is the Riemann tensor of
the connection $D_a$, which we will denote henceforth simply as
$R_{abc}^{\ \ \ d}$.  From the symmetries of the Riemann tensor it
follows that
\begin{equation}
\label{Trautman}
g^{f[a}(\varepsilon)R_{f(bc)}^{\ \ \ \ \ d]}[\nabla_e (\varepsilon)]=0,
\end{equation}
which is called the Trautman condition \cite{Trautman}
\footnote{
For many calculations it is advantageous to rewrite the Trautman
condition as $g^{ea} R_{ebc}^{\ \ \ d} - g^{ed} R_{ecb}^{\ \ \ a} =0$.}.
Inserting the expansion (\ref{metric-up}) of the contravariant
metric and (\ref{Riemann}) of the Riemann tensor, and expanding order
by order in $\varepsilon$, gives the Newton-Cartan Trautman condition
(\ref{Trautman0}) at leading order, and the corresponding
post-Newton-Cartan
condition (\ref{Trautman2}) at subleading order.

\subsubsection{Stress-energy conservation}

Next, we insert the expansions (\ref{Da-def}) and
(\ref{matter_assumpt}) of the
connection and stress-energy tensor into the conservation equation
\begin{equation}
\label{matter-eq}
\nabla_a (\varepsilon)   T^{ab}(\varepsilon) = 0  .
\end{equation}
At leading order this yields the Newtonian stress-energy conservation
equation (\ref{matter-eq_4}), and at subleading order the
post-Newtonian equation (\ref{matter-eq_6}).

\subsubsection{Field equation}

Finally, we write the Einstein field equation (\ref{field-eq00})
in the form
\begin{eqnarray}
\label{fieldeq}
R_{acb}^{\ \ \ \,c}[\nabla_e (\varepsilon)]&=&
4 \pi \left[ 2 g_{ac}(\varepsilon) g_{bd}(\varepsilon) -
  g_{ab}(\varepsilon) g_{cd}(\varepsilon) \right]
T^{cd}(\varepsilon). \nonumber \\
\end{eqnarray}
Inserting the expansions (\ref{Riemann}), (\ref{metric-down}) and
(\ref{matter_assumpt}) of the
Riemann tensor, covariant metric and stress-energy tensor yields
at leading order the Newton-Cartan field equation (\ref{field-eq_0}),
and at subleading order the post-Newton-Cartan field equation
(\ref{field-eq_2}).

\subsection{Combined post-Newton-Cartan theory}
\label{CPNC-derive}

In this section we derive the equations (\ref{orthog0CPNC}) --
(\ref{matter-eq_4CPNC})
of combined post-Newton-Cartan theory from the Newton-Cartan and
perturbative post-Newton-Cartan theories.

Suppose we have a solution of the Newton-Cartan and perturbative
post-Newton-Cartan theories, consisting of the fields $t_a$, $p_{ab}$,
$h^{ab}$, $k^{ab}$, $D_a$, $\Delta^a_{bc}$, ${\cal T}^{ab}$ and ${\cal
  S}^{ab}$.  Such solutions posess a scaling symmetry corresponding
to a change in units of time.  Specifically, it is easy to check that
for any real number $\lambda$ there is a mapping of solutions to
solutions given by rescaling the fields by
$t_a \to e^\lambda t_a$, $k^{ab} \to e^{-2 \lambda} k^{ab}$, ${\cal
  T}^{ab} \to e^{-4 \lambda} {\cal T}^{ab}$, ${\cal S}^{ab} \to e^{-6
  \lambda} {\cal S}^{ab}$, $\Delta^a_{bc} \to e^{-2 \lambda}
\Delta^a_{bc}$, with the other fields being left unchanged.
If we now consider the expansions (\ref{metric-up}), (\ref{metric-down}), (\ref{Da-def}) and
(\ref{matter_assumpt}) of the contravariant metric, covariant metric, connection and
stress-energy tensor, truncated to post-1-Newtonian order, we can
apply this rescaling to effectively set $\varepsilon=1$ in these expansions.
Specifically, for any $\varepsilon>0$, we define the hatted
fields by
\bes
\label{hats}
\bea
{\hat t}_a &=& \frac{1}{\sqrt{\varepsilon}} t_a, \\
{\hat h}^{ab} &=& h^{ab}, \\
{\hat p}_{ab} &=& p_{ab}, \\
{\hat k}^{ab} &=& \varepsilon k^{ab}, \\
{\hat D}_a w_b &=& D_a w_b - \varepsilon \Delta^c_{ab} w_c, \\
{\hat {\cal T}}^{ab} &=& \varepsilon^2 {\cal T}^{ab} + \varepsilon^3 {\cal
  S}^{ab},
\eea
\ees
for any one-form $w_a$.  These will be the fundamental variables of
the combined theory.

Next, we note that the orthogonality conditions (\ref{orthogonality}) and (\ref{inv-metric0})
are preserved under rescaling, which yields the orthogonality
conditions (\ref{orthog0CPNC}) and (\ref{orthogonalityCPNC}) for the
hatted variables.

Next, using the definitions (\ref{hats}) of the hatted fields and the
connection-compatibility conditions
(\ref{compat}), (\ref{compat-upper_2}) and
(\ref{compat-lower_0}) we obtain
\bes
\bea
{\hat D}_a ( {\hat h}^{bc} + {\hat k}^{bc})&=& -\varepsilon^2 (\Delta^{b}_{\ ad}k^{dc}
           -\Delta^{c}_{\ ad}k^{bd}), \ \ \ \\
{\hat D}_a (- {\hat t}_b {\hat t}_c + {\hat p}_{bc})&=&
\varepsilon (\Delta^d_{ab} p_{dc} + \Delta^d_{ac} p_{bd}).
\eea
\ees
The right hand sides of both of these equations are of
post-2-Newtonian order.  Therefore we can drop the right hand sides to
obtain the compatibility conditions (\ref{compatCPNC});
this modification affects the theory only at post-2-Newtonian and
higher orders, and not at Newtonian or post-1-Newtonian order.

Finally, we can use exactly analogous arguments to derive the
Trautman condition (\ref{Trautman0CPNC}), field equation (\ref{field-eq_0CPNC}) and stress-energy
conservation equation (\ref{matter-eq_4CPNC}) of the combined
post-Newton-Cartan theory from
the corresponding equations of the Newton-Cartan and perturbative
post-Newton-Cartan theories.

\section{Derivation of standard, coordinate-specific post-Newtonian
  theory from perturbative post-Newton-Cartan theory}
\label{coord-PN}

\subsection{Change of viewpoint}
\label{shift-view}

So far, we have shown that the Newton-Cartan and post-Newton-Cartan
theories can be derived from general relativity together with the
three assumptions discussed in Sec.\ \ref{sec:foundations}.

We now make a change of viewpoint, and consider these theories as
independent theories in their own right, independent of general
relativity.  In other words, we forget about the spacetime metric,
and instead regard the fields $t_a$, $h^{ab}$, $D_a$ of the Newton-Cartan
theory, and $p_{ab}$, $k^{ab}$ and $\Delta^{c}_{ab}$ of the
post-Newton-Cartan theory, as fundamental.
It is well known that the usual coordinate-dependent formulation of
Newtonian gravity can be derived from the resulting Newton-Cartan theory, under
the assumption of asymptotic flatness.  In this section we will show
that, similarly, the usual coordinate-dependent formulations of
post-1-Newtonian theory can be derived from the post-Newton-Cartan
theory, in suitably chosen coordinate systems, again under the
assumption of asymptotic flatness.

\subsection{Derivation of Newtonian theory from Newton-Cartan theory}
\label{subsec-Newt-coord}

We start by reviewing the well-known derivation at Newtonian order
\cite{Dautcourt}.  We assume that the Newton-Cartan equations
(\ref{orthogonality}) -- (\ref{matter-eq_4})
are valid throughout all of spacetime, and that the Riemann tensor
of the connection $D_a$ goes to zero at spatial infinity (asymptotic
flatness).

From the metric compatibility condition
(\ref{compat-lower_-2}), it follows that there exists a function $t$ on
spacetime for which $t_a = D_a t$.  We will call this function the
time function.  We now introduce a coordinate system $x^\alpha = (x^0,x^j)$
with $x^0 = t.$  The orthogonality condition (\ref{orthogonality}) and
Eq.\ (\ref{compat-lower_-2})
then immediately lead to
\begin{equation}
\label{t-in-coord}
t_{\mu}= t_{,\mu} = \delta_{\mu}^0
\end{equation}
and
\be
\label{ha0}
h^{\mu 0} = 0.
\ee

Next, the compatibility condition (\ref{compat-lower_-2}) gives
$
\partial_{\mu} t_{\nu}
 - \Gamma^{\lambda}_{\mu \nu} t_{\lambda} =0 ,
$
where $\Gamma^{\lambda}_{\mu\nu}$ are the coefficients of the
connection $D_a$,
which yields
\begin{equation}
\Gamma^{0}_{\mu \nu} =0
\label{Gamma0vanish}
\end{equation}
from Eq.\ (\ref{t-in-coord}).
Similarly
the compatibility condition (\ref{compat-upper_0}) yields
\begin{equation}
\label{compat-upper-coord}
\partial_{\mu} h^{\alpha\beta}+\Gamma^{\alpha}_{\mu \nu} h^{\nu\beta}
 +\Gamma^{\beta}_{\mu \nu} h^{\alpha\nu} = 0 .
\end{equation}
If we now define a spatial covariant metric $h_{k j}$ by
\begin{equation}
\label{def-h_kj}
h^{i k}h_{k j} = \delta^{i}_j,
\end{equation}
then Eq.\ (\ref{compat-upper-coord}) results in
\begin{equation}
\Gamma^{i}_{l m}=\frac{1}{2} h^{ik}
\left( h_{kl,m}+h_{km,l}-h_{lm,k} \right),
\label{spatialGamma}
\end{equation}
\begin{equation}
\Gamma^{i}_{0 l}=\frac{1}{2} h^{ik}
\left( h_{kl,0}-\epsilon_{klm} B_m \right),
\end{equation}
and
\begin{equation}
\Gamma^{i}_{0 0}=h^{ik}\Phi_k  .
\label{Gammai00}
\end{equation}
Here the quantities $\Phi_k$ and $B_m$ are
still undetermined.

Next, the field equations (\ref{field-eq_0}) imply
\bes
\bea
\label{Newtonian-i0}
R_{i0} &=& 0,\\
R_{ij} &=& 0,
\label{Newtonian-3Ricci}
\eea
\ees
where $R_{\alpha\beta} = R_{\alpha\beta}[D_a]$ is the Ricci tensor computed from the Newtonian connection $D_a$.
Computing the spatial components of this Ricci tensor
explicitly, and simplifying using the condition (\ref{Gamma0vanish}),
gives
\be
R_{ij} = \partial_k \Gamma^k_{ij} - \partial_i \Gamma^k_{kj} +
\Gamma^k_{kl} \Gamma^l_{ij} - \Gamma^l_{ik} \Gamma^k_{jl}.
\ee
Combining this with Eq.\ (\ref{spatialGamma}) now gives
$
R_{ij} = {}^{(3)}R_{ij}[h_{kl}],
$
where the right hand side is the three-dimensional Ricci tensor
computed from the metric $h_{kl}$ at fixed $t$.
Therefore, from Eq.\ (\ref{Newtonian-3Ricci}), the Ricci tensor
of the metric
$h_{ij}$ vanishes.  Since this metric is three dimensional, it follows
that the metric is flat.  Hence at each
fixed $t$ we can choose the spatial coordinates so that
\be
h_{ij} = \delta_{ij}.
\label{flatcoords}
\ee

Next, combining the field equation (\ref{Newtonian-i0}) with the
connection coefficients (\ref{spatialGamma}) -- (\ref{Gammai00})
and simplifying using the coordinate condition (\ref{flatcoords})
gives
\be
\epsilon_{ijk} \partial_j B_k=0.
\label{f1}
\ee
Also the Trautman condition (\ref{Trautman0}) implies that ${\bf B}$
is transverse, $\partial_i B_i=0$.
Together with Eq.\ (\ref{f1}) this implies that the field ${\bf B}$ satisfies
Laplace's equation,
$B_{i,jj}=0$.  Next, using the assumption that
$R_{\alpha\beta\gamma}^{\ \ \ \ \delta} \to 0$ as $r \to \infty$, we
find that the only allowed nontrivial solutions to Laplace's equation
are those with ${\bf B} =$ constant.
These solutions
can be eliminated by transforming to a uniformly rotating coordinate
system.  It
follows that, in a suitably adjusted coordinate system, $B_i=0$.
Finally, the Trautman condition (\ref{Trautman0}) implies that
$
\Phi_{i,j} - \Phi_{j,i} = \epsilon_{ijk} {\dot B}_{k} =0,
$
so that $\Phi_i = \partial_i \Phi$
for some function $\Phi$, which will be the Newtonian potential.  It
follows that
\be
\Gamma^i_{00} = \Phi_{,i},
\label{Gammai00v}
\ee
while all the other connection coefficients vanish.
The field equation (\ref{field-eq_0}) then reduces to the Poisson
equation $\Phi_{,kk} = 4 \pi {\cal T}^{00}$, and the stress-energy
conservation equation (\ref{matter-eq_4}) reduces to
$\partial_\alpha {\cal T}^{\alpha0}=0$, $\partial_\alpha {\cal
  T}^{\alpha i} + \Phi_{,i} {\cal T}^{00} =0$.  These are the standard
equations of Newtonian gravity.

We note that the Newton-Cartan theory contains more local degrees of
freedom that Newtonian theory.  In particular, if one assumes that
Newton-Cartan theory holds only in a local region of spacetime, or if
one assumes it holds everywhere but drops the assumption of asymptotic
flatness, then one obtains a theory with an additional transverse
vector field ${\bf B}$ that satisfies Laplace's equation. This is just the
gravitomagnetic field which normally arises at post-1-Newtonian order.
Thus, the Newton-Cartan theory admits source-free gravitomagnetic fields at
Newtonian order.

Another method that has been used in the literature to exclude these
extra degrees of freedom is to assume that \cite{Ehlers-Rahmen,Ehlers-frame,MTW}
\be
h^{ec}R_{abc}^{\ \ \  \ d}=0,
\label{flatt}
\ee
where $R_{abc}^{\ \ \ \ d}$ is the Riemann tensor of the
connection $D_a$.
Augmenting the equations of Newton-Cartan theory
with this assumption
yields
a covariant theory which is equivalent, locally, to Newtonian gravity.
However, the assumption (\ref{flatt}) cannot be derived from general
relativity in a local manner; its validity requires the use of global
information.  For this reason we do not use the assumption (\ref{flatt})
in this paper.

\subsection{Derivation of post-Newtonian theory from
perturbative  post-Newton-Cartan theory}
\label{sec-PN-fields}

We now extend the above derivation to post-1-Newtonian order.
We continue to use the adapted coordinate system derived above,
and we assume that the post-Newton-Cartan equations
are valid throughout all of spacetime.

The spatial components of the orthogonality condition
(\ref{inv-metric0}) imply that
$
h^{i k}p_{k j} = \delta^{i}_j  .
$
Hence the covariant spatial metric $h_{kj}$ defined in Eq.\
(\ref{def-h_kj}) and the spatial components of $p_{ab}$ coincide,
and in our adapted coordinate system we have
\begin{equation}
p_{i j} = h_{ij} = \delta_{ij} .
\label{pij}
\end{equation}
The remaining components of the orthogonality condition
(\ref{inv-metric0}) yield
\begin{equation}
k^{00}= -1  ,
\label{k00}
\end{equation}
and
\begin{equation}
k^{i0} = h^{i k}p_{k 0} = p_{i0} .
\label{ki0}
\end{equation}

In order to find expressions for $p_{i0}$ and $p_{00}$ we
have to consider the metric compatibility condition
(\ref{compat-lower_0}), which yields
\begin{equation}
2\Delta^{0}_{\ \mu 0}=-p_{00,\mu}
+2\Delta^{\lambda}_{\mu 0}p_{\lambda 0}.
\end{equation}
It follows that
\begin{equation}
\label{Delta_0_00_I}
\Delta^{0}_{\ 0 0}=-\frac{1}{2}p_{00,0}+\Phi_{,l}p_{l0}
\end{equation}
and
\begin{equation}
\label{Delta_0_j0_I}
\Delta^{0}_{\ j 0}=-\frac{1}{2}p_{00,j}.
\end{equation}
The condition
(\ref{compat-lower_0}) also yields
\begin{equation}
\Delta^{0}_{\ \mu j}=-p_{0j,\mu}
+\Delta^{\lambda}_{\mu j}p_{\lambda 0}
+\Delta^{\lambda}_{\mu 0}p_{j \lambda} ,
\end{equation}
which implies
\begin{equation}
\label{Delta_0_ij_I}
\Delta^{0}_{\ i j}=-p_{0j,i}
\end{equation}
and
\begin{equation}
\label{Delta_0_0j_I}
\Delta^{0}_{\ 0 j}=\Phi_{,j}-p_{j0,0} .
\end{equation}
Since the connection is symmetric, Eq.\ (\ref{Delta_0_ij_I}) yields
$
p_{0i,j}-p_{0j,i}=0 ,
$
and thus
$
p_{0i}= g_{,i} ,
$
for some function $g$.
Now under infinitesimal gauge transformations $p_{0i}$ transforms as, from Eq.\ (\ref{gamma-trafo}),
\begin{equation}
\label{gauge-trafo-of-gamma_0i}
p_{0i} \rightarrow p_{0i} - \xi^{0}_{\ ,i} ,
\end{equation}
where $\xi^{0}$ is an
arbitrary function.  Therefore by taking $\xi^0=g$ we can specialize
the gauge to enforce
\begin{equation}
\label{gamma_i0}
p_{0i} = 0  .
\end{equation}

Next, the symmetry of the connection applied to Eqs.\
(\ref{Delta_0_j0_I}) and
(\ref{Delta_0_0j_I}) yields together with Eq.\ (\ref{gamma_i0}) that
$
p_{00,j} = -2\Phi_{,j}  ,
$
which implies
$
p_{00} = -2\Phi +\chi(x^0 )  ,
$
where $\chi(x^0 )$ is a function which depends only on $x^0$.
Now note that the gauge condition (\ref{gamma_i0}) does not completely
fix the gauge; from Eq.\ (\ref{gauge-trafo-of-gamma_0i})
gauge transformations with $\xi^{0}_{\ ,i}=0$ leave the condition
(\ref{gamma_i0}) invariant.
Since $p_{00}$ transforms like
$
p_{00} \rightarrow p_{00} - 2\xi^{0}_{\ ,0}
$
under the infinitesimal gauge transformations of Eq.\ (\ref{gamma-trafo}),
we can further specialize the gauge to enforce
\begin{equation}
\label{gamma_00}
p_{00} = -2\Phi .
\end{equation}
Simplifying Eqs.
(\ref{Delta_0_00_I}), (\ref{Delta_0_j0_I}) and (\ref{Delta_0_ij_I})
using the gauge conditions
Eqs.\ (\ref{gamma_i0}) and (\ref{gamma_00}) now
yields
\begin{equation}
\Delta^{0}_{\ 0 0}= \Phi_{,0}, \ \ \
\Delta^{0}_{\ i 0}=\Phi_{,i}, \ \ \
\Delta^{0}_{\ i j}=0.
\label{Delta0ab}
\end{equation}

Next we determine the remaining components of
$\Delta^{c}_{\ ab}$.
From the compatibility condition (\ref{compat-upper_2}) we get
\begin{equation}
\Delta^{i}_{\ \alpha l}h^{lj} + \Delta^{j}_{\ \alpha l}h^{li}
=-k^{ij}_{\ \ ,\alpha}-\Gamma^{i}_{\alpha \lambda}k^{\lambda j}
-\Gamma^{j}_{\alpha \lambda}k^{i \lambda }.
\end{equation}
This leads to
\begin{equation}
\Delta^{i}_{\ k j} =-\frac{1}{2}k^{ij}_{\ \ ,k}+W_{ijk}
\label{124}
\end{equation}
and
\begin{equation}
\label{Delta_i_0j-I}
\Delta^{i}_{\ 0 j} =-\frac{1}{2}k^{ij}_{\ \ ,0}+V_{ij}  ,
\end{equation}
where $W_{ijk}=-W_{jik}$ and $V_{ij}=-V_{ji}$ are undetermined.
Since the connection is symmetric it follows from Eq.\ (\ref{124}) that
\begin{equation}
W_{ijk}=W_{ikj}+\frac{1}{2}\left(k^{ij}_{\ \ ,k}-k^{ik}_{\ \ ,j}\right) .
\end{equation}
Together with $W_{ijk}=-W_{jik}$ this implies that
\begin{equation}
\label{Delta_i_kj}
\Delta^{i}_{\ jk} =
\frac{1}{2}\left(k^{jk}_{\ \ ,i}-k^{ij}_{\ \ ,k}-k^{ik}_{\ \ ,j}\right) .
\end{equation}

Thus, from the metric compatibility conditions
(\ref{compat-lower_0}) and (\ref{compat-upper_2})
we have been able to determine all the components of $\Delta^{\lambda
 }_{\ \mu\nu}$, except for $\Delta^{i}_{\ 0 j}$ and $\Delta^{i}_{\ 0
   0}$.  To determine those components we use the
Trautman condition (\ref{Trautman2}),
which can be rewritten as
\begin{eqnarray}
\label{Trau2-upper2}
k^{de}R_{ecb}^{\ \ \  a}-k^{ae}R_{ebc}^{\ \ \  d}&=&
\left( D_b \Delta^{d}_{\ ec}- D_e \Delta^{d}_{\ bc}\right)h^{ae}
\nonumber \\
&+&\left( D_e \Delta^{a}_{\ bc}- D_c \Delta^{a}_{\ be} \right)h^{de}.
\nonumber \\
\end{eqnarray}
Specializing to
$a=n$, $b= 0$, $c= k$ and $d=m$
gives
\begin{equation}
  D_0 \Delta^{m}_{\ nk}- D_n \Delta^{m}_{\ 0k}
+ D_m \Delta^{n}_{\ 0k}- D_c \Delta^{n}_{\ 0m}
=0,
\end{equation}
which using Eqs.\ (\ref{Delta_i_kj}) and  (\ref{Delta_i_0j-I})
simplifies to
\begin{equation}
\label{eq_for_V}
V_{nk,m} - V_{mk,n} - V_{nm,k} = 0 .
\end{equation}
Since $V_{nm}$ is antisymmetric the solution of (\ref{eq_for_V})
is
\begin{equation}
\label{V_mn}
V_{mn}=\frac{1}{2}\left( \gamma_{m,n} - \gamma_{n,m} \right) ,
\end{equation}
where $\gamma_{m}$ is some undetermined vector field.

Next, considering the $a\to n$, $b=c\to 0$, $d\to m$
components of Eq.\ (\ref{Trau2-upper2}),
we find
\begin{eqnarray}
\label{Delta_m_00-I}
D_0 \Delta^{m}_{\ n0}- D_n \Delta^{m}_{\ 00}
+D_m \Delta^{n}_{\ 00}- D_0 \Delta^{n}_{\ 0m}
\nonumber \\
=\frac{1}{2} \left(k^{ml} p_{00,nl}-k^{nl} p_{00,ml} \right).
\end{eqnarray}
If we define
\begin{equation}
\label{A_m-def}
A_m \equiv \Delta^{m}_{\ 00}+ \frac{1}{2} p_{00,l}k^{ml}
      -\gamma_{m,0},
\end{equation}
then Eq.\ (\ref{Delta_m_00-I}) becomes
\begin{equation}
A_{m,n} - A_{n,m} = 0  ,
\end{equation}
so we can write
\begin{equation}
\label{A_m-soln}
A_{m}=-\frac{1}{2}\gamma_{,m}   ,
\end{equation}
for some function $\gamma$.
Combining Eqs.\ (\ref{A_m-def}) and (\ref{A_m-soln})
now yields
\begin{equation}
\label{Delta_i_00}
\Delta^{i}_{\ 00}= -\frac{1}{2}\gamma_{,i}+\gamma_{i,0}
-\frac{1}{2}p_{00,l}k^{li}  ,
\end{equation}
while Eqs.\ (\ref{V_mn}) and (\ref{Delta_i_0j-I}) lead to
\begin{equation}
\label{Delta_i_0j}
\Delta^{i}_{\ 0j}=
\frac{1}{2}\left(\gamma_{i,j}-\gamma_{j,i}\right)
-\frac{1}{2}k^{ij}_{\ \ ,0}  .
\end{equation}

We now define a metric ${\hat g}_{ab}(\varepsilon)$, for each
$\varepsilon>0$, by the formula
\begin{equation}
\label{metric-down1}
{\hat g}_{ab}(\varepsilon)=- \frac{1}{\varepsilon} t_a t_b + p_{ab}
+ \varepsilon q_{ab},
\end{equation}
cf.\ the metric expansion (\ref{metric-down}) above.
Here the tensor $q_{ab}$ is defined by
\bes
\label{qabvalues}
\bea
q_{00} &=& \gamma, \\
q_{0i} &=& \gamma_i, \\
q_{ij} &=& - k^{ij}.
\label{qijv}
\eea
\ees
We compute the inverse of this metric, using the values of the components of
the fields $t_a$, $p_{ab}$ and $q_{ab}$ given in Eqs.\ (\ref{t-in-coord}),
(\ref{pij}), (\ref{gamma_i0}), (\ref{gamma_00}) and (\ref{qabvalues}).
The result is of the form [cf.\ Eq.\ (\ref{metric-up}) above]
\begin{equation}
\label{metric-up1}
{\hat g}^{ab}(\varepsilon)=h^{ab}+\varepsilon k^{ab} + \varepsilon^2 j^{ab} +
O(\varepsilon^3).
\end{equation}
Here the components of the fields $h^{ab}$ and $k^{ab}$ (except for
$k^{ij}$) are those given
by Eqs.\ (\ref{ha0}), (\ref{flatcoords}), (\ref{k00}), (\ref{ki0}) and
(\ref{gamma_i0}).  This result is guaranteed because we have imposed
the orthogonality conditions (\ref{orthogonality}) and
(\ref{inv-metric0}).  The fact that the spatial components of the
coefficient of $\varepsilon$ in Eq.\ (\ref{metric-up1}) are $k^{ij}$
follows from
the choice (\ref{qijv}) of spatial components of $q_{ab}$.
The explicit form of the tensor $j^{ab}$ which appears in Eq.\
(\ref{metric-up1}) will not be needed in what follows.

Next, we compute the coefficients ${\hat
  \Gamma}^\alpha_{\beta\gamma}(\varepsilon)$ of the Levi-Civita
connection associated with the metric
(\ref{metric-down1}), using the expansion (\ref{metric-up1}).
Suppressing indices, the result is schematically of the form
\begin{eqnarray}
{\hat \Gamma} &\sim& \frac{1}{\varepsilon} (h  t \partial t) +
(h \partial p + k t \partial t) \nonumber \\
&&+ \varepsilon (h \partial q +
k \partial p + j t \partial t) + O(\varepsilon^2).
\label{LC}
\end{eqnarray}
The leading order, $O(\varepsilon^{-1})$ term vanishes identically by
virtue of Eq.\ (\ref{t-in-coord}).
We can evaluate the
next order, $O(\varepsilon^{0})$ term using the
the specific values of the components of $t_a$, $p_{ab}$, $h^{ab}$ and
$k^{ab}$ in our adapted coordinate system, given by Eqs.\
(\ref{t-in-coord}), (\ref{ha0}), (\ref{flatcoords}), (\ref{pij}),
(\ref{k00}), (\ref{ki0}), (\ref{gamma_i0}) and (\ref{gamma_00}).
The resulting expressions are just the coefficients
$\Gamma^\alpha_{\beta\gamma}$ of the Newton-Cartan connection $D_a$, with
the only nonzero component being given by Eq.\ (\ref{Gammai00v}).
Again, this result is not surprising because we have enforced the
compatibility conditions (\ref{compat-upper_0}),
(\ref{compat-lower_-2}), (\ref{compat-upper_2}) and (\ref{compat-lower_0}).
Similarly, at the next order, we find that the $O(\varepsilon)$
components in Eq.\ (\ref{LC}) coincide with the components of the
post-Newton-Cartan field $\Delta^\alpha_{\beta\gamma}$, given by Eqs.\
(\ref{Delta0ab}), (\ref{Delta_i_kj}), (\ref{Delta_i_00}) and
(\ref{Delta_i_0j}). [Note that the result is independent of
$j^{\alpha\beta}$, by virtue of Eq.\ (\ref{t-in-coord})].

To summarize, we have been able to show that all of our Newton-Cartan
and post-Newton-Cartan fields can be derived from the three fields
$t_a$, $p_{ab}$ and $q_{ab}$ that enter into the expansion (\ref{metric-down1}) of
the metric, using the standard equations of general relativity.
Moreover, that metric expansion is of the standard post-Newtonian form;
using the specific values of the components of $t_a$, $p_{ab}$ and
$q_{ab}$ given above and writing $\varepsilon = c^{-2}$, Eq.\
(\ref{metric-down1}) takes the form
\bea
ds^2 &=& - c^2 \left[ 1 + \frac{2 \Phi}{c^2} - \frac{\gamma}{c^4} +
  O\left(\frac{1}{c^6} \right) \right] dt^2
\nonumber \\ &&
+ 2 \left[ \frac{\gamma_i}{c^2} + O\left( \frac{1}{c^4} \right)
\right] dx^i dt
\nonumber \\ &&
+ \left[ \delta_{ij} + \frac{1}{c^2} q_{ij} + O\left(
    \frac{1}{c^4} \right) \right] dx^i dx^j.
\eea
This is the standard starting point for coordinate-specific
post-Newtonian theory, involving a post-Newtonian correction to the
Newtonian potential $\Phi$, and a gravitomagnetic potential
$\gamma_i$ \footnote{The spatial tensor $q_{ij}$ does not contain any
independent degrees of freedom; using the post-Newtonian field
equations and making a gauge
specialization gives $q_{ij} = -2 \Phi \delta_{ij}$ \protect{\cite{Damour}}.  This gauge
specialization is nearly always adopted in post-Newtonian theory.}.
It follows that all of the relations of the Newton-Cartan and
post-Newton-Cartan theories, when expressed in terms $\Phi$, $\gamma$,
$\gamma_i$ and $q_{ij}$,
are either identically satisfied, or
reduce to the Einstein equations that one would compute directly from
the metric (\ref{metric-down1}), i.e. the coordinate-specific
post-Newtonian equations.
Furthermore we have shown how to obtain the quantities $\Phi$,
$\gamma$, $\gamma_i$ and $q_{ij}$, starting from solutions of the
Newton-Cartan and post-Newton-Cartan equations.
It follows that the equations of the
perturbative post-Newton-Cartan theory are equivalent to those of the standard
coordinate-specific post-Newtonian theory.

\section{Discussion and Conclusions}
\label{Conclusion}

We have derived a covariant version of the equations of the
post-1-Newtonian approximation to general relativity.
These equations reduce to the standard coordinate
formulation of post-Newtonian theory in asymptotically flat spacetimes
in suitable coordinate systems.

Although the covariant formulation is elegant,
it does not provide a very
compact or efficient representation of the theory.  In a
general coordinate system the combined post-Newton-Cartan theory
involves 74 free functions to describe the geometry, as compared to
four for standard post-Newtonian theory.  The covariant formulation is
therefore mostly of formal interest.  However, it may be useful
to connect the different gauge-dependent formulations that are found
in the literature.  It may also be useful for deriving general
properties of post-Newtonian theory.
It might also provide insight into the meaning of the parameters of the
parametrized post-Newtonian framework \cite{Will-book}, if the
analysis of this
paper were generalized to the class of theories of gravity encompassed
by that framework.

\acknowledgments

This research was supported in part by NSF grants PHY-9722189,
PHY-0757735 and PHY-0855315.


\appendix

\section{Limiting behavior of connection derived from other postulates}
\label{ThirdAssumption}

In this Appendix we show that our assumption 3 on the limiting behavior of the connection will hold (up to gauge transformations) whenever
the Riemann tensor is bounded as $\varepsilon \to 0$.  More precisely,
whenever assumptions 1 and 2 hold
in a local region, and the Riemann tensor $R_{abc}^{\ \ \ \
  d}(\varepsilon)$ is finite as $\varepsilon \to 0$, then
we show that there exists a (possibly irregular) gauge
transformation of the form (\ref{general-gauge-transformation}) such
that the transformed one-parameter family of
solutions satisfies assumptions 1, 2 and 3 of Sec.\
\ref{sec:assumptions}.

We start by fixing a coordinate system and computing the
connection coefficients using the expansions (\ref{metric-up}) and (\ref{metric-down}) of
the contravariant and covariant metrics and the orthogonality
condition (\ref{orthogonality}).  The result is of the form
[cf.\ Eq.\ (\ref{Gamma-schematic}) above]
\be
\Gamma^\alpha_{\beta\gamma}(\varepsilon) =
\varepsilon^{-1}
\Gamma^{(-1)\alpha}_{\ \ \ \ \, \beta\gamma}
+ \Gamma^{(0)\alpha}_{\ \ \ \beta\gamma}
+ \varepsilon \Gamma^{(1)\alpha}_{\ \ \ \beta\gamma} + O(\varepsilon^2),
\ee
with
\be
\Gamma^{(-1)\alpha}_{\ \ \ \ \, \beta\gamma} = - h^{\alpha\lambda} (
t_\gamma t_{[\lambda,\beta]} + t_\beta t_{[\lambda,\gamma]} )
\label{Gamma-1}
\ee
and
\bea
\Gamma^{(0)\alpha}_{\ \ \ \beta\gamma} &=& - k^{\alpha\lambda} (
t_\gamma t_{[\lambda,\beta]} + t_\beta t_{[\lambda,\gamma]} +
t_\lambda t_{(\beta,\gamma)}) \nonumber \\
&& + \frac{1}{2} h^{\alpha\lambda} (- p_{\beta\gamma,\lambda} +
p_{\beta\lambda,\gamma} + p_{\gamma\lambda,\beta} ).
\label{Gamma0}
\eea
Here $\Gamma^{(0)\alpha}_{\ \ \ \beta\gamma}$ are the
coefficients of the Newtonian connection $D_a$, which were denoted
simply $\Gamma^\alpha_{\beta\gamma}$ in the body of the paper.  Also
$\Gamma^{(1)\alpha}_{\ \ \ \beta\gamma}$ are the
coefficients of the post-Newtonian connection perturbation, which were
denoted $\Delta^\alpha_{\beta\gamma}$ in the body of the paper.
We want to show that $\Gamma^{(-1)\alpha}_{\ \ \ \ \, \beta\gamma}$ vanishes.

We next compute the expansion of the Einstein tensor, which is of the
form
\bea
G^{\alpha\beta}(\varepsilon) &=&
\varepsilon^{-2} G^{(-2)\alpha\beta}
+ \varepsilon^{-1} G^{(-1)\alpha\beta}
+ G^{(0)\alpha\beta} \nonumber \\
&& + \varepsilon G^{(1)\alpha\beta} + O(\varepsilon^2).
\eea
It follows from assumption 2 of Sec.\ \ref{sec:assumptions} that
the first four terms in this expansion all vanish.  We find that
$G^{(-2)\alpha\beta}$ vanishes identically, while
$G^{(-1)\alpha\beta}$ is given by
\be
G^{(-1)\alpha\beta} = 2 H^\alpha_{\ \, \mu} H^\beta_{\ \, \nu} h^{\mu\nu} +
\frac{1}{2} h^{\alpha\beta} H^\mu_{\ \, \nu} H^\nu_{\ \, \mu},
\label{Einstein-1}
\ee
where $H^\alpha_{\ \, \beta} = h^{\alpha\gamma} t_{[\gamma,\beta]}$.
Setting this expression to zero yields $H^\mu_{\ \, \nu} H^\nu_{\ \,
  \mu}=0$, from which it follows \cite{Dautcourt} that
\be
t_\alpha = f t_{,\alpha}
\ee
for some functions $f$ and $t$.  We now specialize the coordinates
by choosing $x^0 = t$, so that $h^{0\alpha} =0$ and $t_\alpha = f
\delta^0_\alpha$.

We now extend this computation to the next order.  The orthogonality
relation (\ref{inv-metric0}) implies that
\be
k^{00} = -1/f^2, \ \ \ k^{0i} = h^{ij} p_{0j}/f^2, \ \ \ h^{ij} p_{jk}
= \delta^i_k.
\ee
Using these relations and the expansions (\ref{Gamma-1}) and
(\ref{Gamma0}) gives $G^{(0)00} = G^{(0)0i} = 0$ and
\be
G^{(0)ij} = G^{ij}[h_{kl}]  -
\frac{1}{f} D^i D^j f + \frac{1}{2f} h^{ij} D^k D_k f.
\label{Einstein0}
\ee
Here the first term denotes the three-dimensional Einstein tensor
computed from the metric $h_{ij} = p_{ij}$ (the inverse of $h^{ij}$),
and $D_i$ is the covariant derivative associated with that metric.
Also the $O(\varepsilon^{-1})$ piece of the Riemann tensor is given by
\be
R^{(-1)\ \ j}_{\ \ \ \ 0i0} = f D_i D^j f,
\label{Riemann-1}
\ee
with the other components being zero.
Our assumption on the Riemann tensor forces this quantity to vanish,
from which it follows from the vanishing of the expression
(\ref{Einstein0}) for $G^{(0)ij}$ that the Einstein tensor of the
metric $h_{ij}$ must be zero.  We can therefore specialize the
coordinates so that $h_{ij} = h^{ij} = \delta_{ij}$.

It now follows from Eq.\ (\ref{Riemann-1}) that $f_{,ij} =0$, so
that $f = \alpha(t) + \beta_i(t) x^i$.  The leading order expression
for the metric is therefore the Rindler metric, and we can apply the
standard gauge transformation\footnote{In defining this gauge
  transformation we treat $\alpha$ and $\beta_i$ as constants; their
  time dependence affects the final metric only at subleading order.
The gauge transformation depends explicitly on $\varepsilon$ and is
not smooth as $\varepsilon \to 0$.}
that takes the Rindler metric to the
Minkowski metric.  The result of this transformation is to
effectively set $f$ to unity, and so from Eq.\ (\ref{Gamma-1}) it
follows
that $\Gamma^{(-1)\alpha}_{\ \ \ \ \, \beta\gamma} =0$ for the
transformed one-parameter family of metrics.

\section{Gauge freedom in the post-Newtonian fields}
\label{subsec-gauge-cov}

In this Appendix we derive how the Newtonian and post-Newtonian fields
transform under a regular gauge transformation $\varphi_\varepsilon$ of the form
(\ref{general-transf}).  Such a gauge transformation is
parameterized by an
$\varepsilon$-independent
diffeomorphism $\varphi_0$, and by a set of vector fields ${\vec
  \xi}_{(1)}, {\vec \xi}_{(2)}, \ldots$, one for each order in
$\varepsilon$.  For simplicity, we will take $\varphi_0$ to be the
identity mapping, since all quantities will transform trivially under
this portion of the overall diffeomorphism.
Consider now any tensor field $S(\varepsilon)$ which depends on
$\varepsilon$, and has an expansion of the form
\be
S(\varepsilon) = S^{(0)} + \varepsilon S^{(1)} + \varepsilon^2 S^{(2)}
+ O(\varepsilon^3).
\ee
Here for brevity we have suppressed any tensor indices on $S$.
We define the transformed expansion coefficients ${\bar S}^{(j)}$ via
the expansion
\be
\varphi_{\varepsilon\,*} S(\varepsilon) = {\bar S}(\varepsilon) = {\bar
S}^{(0)} + \varepsilon {\bar S}^{(1)} + \varepsilon^2 {\bar S}^{(2)}
+ O(\varepsilon^3).
\ee
From Eq.\ (\ref{general-transf}) it now follows that \cite{FW}
\bes
\label{transfrule}
\bea
{\bar S}^{(0)} &=&  S^{(0)},  \\
{\bar S}^{(1)} &=& S^{(1)} + {\cal L}_{{\vec \xi}_1}
S^{(0)} , \\
{\bar S}^{(2)} &=& S^{(2)}
+ {\cal L}_{{\vec \xi}_2} S^{(0)}
+ {\cal L}_{{\vec \xi}_1} S^{(1)}
\nonumber \\
&&+ {1 \over 2} {\cal L}_{{\vec \xi}_1} {\cal L}_{{\vec \xi}_1}
 S^{(0)},
\eea
\ees
where ${\cal L}$ is the Lie derivative.

We now apply this formalism to the expansions
(\ref{metric-up}), (\ref{metric-down}),
(\ref{matter_assumpt}) and (\ref{connection-limit}) of the contravariant metric,
covariant metric,
stress-energy tensor and connection.  We use of the compatibility
conditions (\ref{compat}), denote gauge-transformed quantities with
bars, and rewrite Lie derivatives in terms of $D_a$ derivatives.
This yields that the Newtonian fields $h^{ab}$, $t_a$, $D_a$ and
${\cal T}^{ab}$ are invariant, while the post-Newtonian fields
transform as
\bes
\label{gts}
\bea
\label{h-trafo}
{\bar k}^{ab} &=&  k^{ab} -h^{ac}D_c \xi^b - h^{bc}D_c \xi^a , \\
\label{gamma-trafo}
{\bar p}_{ab} &=& p_{ab} -t_a t_c D_b \xi^c -t_b t_c D_a \xi^c , \\
\label{Delta-trafo}
{\bar \Delta}^{c}_{\ ab} &=& \Delta^{c}_{\ ab}
                            -2 \xi^d R_{d(ab)}^{\ \ \ \ \ c} + D_{(a} D_{b)} \xi^c, \\
\label{T-trafo}
{\bar {\cal S}}^{ab} &=& {\cal S}^{ab} + \xi^c D_c {\cal T}^{ab}
          -2 {\cal T}^{c(a} D_c \xi^{b)} .
\eea
\ees
Here we have written ${\vec \xi}_{(1)}$ simply as ${\vec \xi}$.
One can check that the post-Newton-Cartan equations
(\ref{inv-metric0})--(\ref{matter-eq_6})
are invariant under these transformations, as they must be.

To obtain the formula (\ref{Delta-trafo}),
let $\omega_b(\varepsilon)$ be an arbitrary one form which depends
smoothly on $\varepsilon$, with the expansion
$\omega_b = \omega_b^{(0)} + \varepsilon \omega_b^{(1)} + O(\varepsilon^2)$.
We define the tensor $S_{ab} = \nabla_a \omega_b$, which has the expansion
\begin{eqnarray}
\label{Sexpandd}
S_{ab}(\varepsilon)
&=& S_{ab}^{(0)} + \varepsilon S_{ab}^{(1)} + O(\varepsilon^2) \\
&=& D_a \omega_b^{(0)} + \varepsilon \left[ D_a \omega_b^{(1)} -
\Delta^{c}_{\ ab} \omega_c^{(0)}  \right]    + O(\varepsilon^2) . \nonumber
\end{eqnarray}
Applying the general transformation rule (\ref{transfrule}) now yields
\begin{equation}
\label{S2-transform}
S_{ab}^{(1)} \rightarrow S_{ab}^{(1)} + {\cal L}_{\xi} S_{ab}^{(0)},
\end{equation}
and
\begin{equation}
\label{w1-transform}
\omega_{a}^{(1)} \rightarrow \omega_{a}^{(1)} + {\cal L}_{\xi} w_{a}^{(0)}.
\end{equation}
Combining Eqs.\ (\ref{Sexpandd}) -- (\ref{w1-transform}) now yields
\begin{eqnarray}
\Delta^{c}_{\ ab}\omega_c^{(0)}
& \rightarrow  &
\Delta^{c}_{\ ab}\omega_c^{(0)}
  + \xi^d D_a D_d \omega_b^{(0)}
  - \xi^d D_d D_a \omega_b^{(0)} \nonumber \\
&&+ \omega_d^{(0)} D_a D_b \xi^d
\end{eqnarray}
which results in Eq.\ (\ref{Delta-trafo}).

\section{perfect fluids}
\label{perfect-fluid}

In this Appendix we describe as an example how perfect fluids can be
described in the covariant formalism.  The stress-energy tensor is
\be
T^{ab} = (\rho+p) u^a u^b + p g^{ab},
\ee
where $\rho$ is the density, $p$ the pressure and $u^a$ the
four-velocity.  The appropriate form of the expansions of these fields
is
\bes
\bea
u^a &=& \sqrt{\varepsilon} \left[ u^a_{\rm n} + \varepsilon u^a_{\rm
    pn}  + O(\varepsilon^2) \right], \\
\rho &=& \varepsilon \rho_{\rm n} + \varepsilon^2 \rho_{\rm pn} +
O(\varepsilon^3), \\
p &=& \varepsilon^2 p_{\rm n} + \varepsilon^3 p_{\rm pn} +
O(\varepsilon^3).
\eea
\ees
Here the subscripts ``n'' and ``pn'' indicate the Newtonian-order and
post-Newtonian order pieces.  Comparing with the expansion
(\ref{matter_assumpt}) of
the stress-energy tensor yields the formulae
\bes
\bea
{\cal T}^{ab} &=& \rho_{\rm n} u_{\rm n}^a u_{\rm n}^b + p_{\rm n}
h^{ab}, \\
{\cal S}^{ab}  &=& (\rho_{\rm pn} + p_{\rm n}) u_{\rm n}^a u_{\rm n}^b
+ 2 \rho_{\rm n} u_{\rm n}^{(a} u_{\rm pn}^{b)}
+ p_{\rm n} k^{ab} \nonumber \\ &&
+ p_{\rm pn} h^{ab}.
\eea
\ees
Also the normalization of the four-velocity yields the conditions
\be
t_a u_{\rm n}^a =1, \ \ \ p_{ab} u_{\rm n}^a u_{\rm n}^b = 2 t_a u_{\rm pn}^a.
\label{normc}
\ee
One can check that inserting these expressions in the stress-energy
conservation laws (\ref{matter-eq_4}) and (\ref{matter-eq_6}), using the
specific forms of $h^{ab}$, $k^{ab}$ and $\Delta^a_{bc}$ derived in
Sec.\ \ref{coord-PN} and using the normalization constraints (\ref{normc}) yields the usual equations of
Newtonian and post-Newtonian hydrodynamics.

Alternatively, one can combine the Newtonian and post-Newtonian pieces
together, as in the combined post-Newton-Cartan theory derived in Sec.\
\ref{CPNC-derive}.  Defining ${\hat \rho} = \rho_{\rm n} + \varepsilon \rho_{\rm pn}$, ${\hat p} = p_{\rm n} + \varepsilon p_{\rm pn}$ and ${\hat u}^a = u^a_{\rm n} + \varepsilon u^a_{\rm pn}$, then the combined stress energy of Sec.\ \ref{CPNC-derive} is
\be
{\hat {\cal T}}^{ab} = ({\hat \rho} + {\hat p}) {\hat u}^a {\hat u}^b + {\hat p} ({\hat h}^{ab} + {\hat k}^{ab}),
\ee
and the normalization constraint is ${\hat u}^a {\hat u}^b ({\hat t}_a {\hat t}_b - {\hat p}_{ab})=-1$.
Again one can check that these expressions lead to the usual post-Newtonian hydrodynamic equations.

\end{document}